\documentclass[sigconf, nonacm]{acmart}
\AtBeginDocument{%
  }

\copyrightyear{2025}
\acmYear{2025}
\setcopyright{acmlicensed}\acmConference[SA Technical Communications '25]{SIGGRAPH Asia 2025 Technical Communications}{December 15--18, 2025}{Hong Kong, Hong Kong}
\acmBooktitle{SIGGRAPH Asia 2025 Technical Communications (SA Technical Communications '25), December 15--18, 2025, Hong Kong, Hong Kong}
\acmDOI{10.1145/3757376.3771403}
\acmISBN{979-8-4007-2136-6/2025/12}

\acmSubmissionID{158}

\begin{document}

\title[SketchRodGS: Sketch-based Extraction of Slender Geometries from 3DGS]{SketchRodGS: Sketch-based Extraction of Slender Geometries for Animating Gaussian Splatting Scenes}

\author{Haato Watanabe}
\affiliation{%
  \institution{The University of Tokyo}
  \country{Japan}
}
\orcid{0009-0006-0097-9310}
\email{heart.watanabe.research@gmail.com}

\author{Nobuyuki Umetani}
\affiliation{%
  \institution{The University of Tokyo}
  \country{Japan}
}
\orcid{0000-0003-1251-970X}
\email{n.umetani@gmail.com}

\renewcommand{\shortauthors}{Watanabe et al.}

\begin{abstract}
Physics simulation of slender elastic objects often requires discretization as a polyline. 
However, constructing a polyline from Gaussian splatting is challenging as Gaussian splatting lacks connectivity information and the configuration of Gaussian primitives contains much noise. 
This paper presents a method to extract a polyline representation of the slender part of the objects in a Gaussian splatting scene from the user's sketching input. 
Our method robustly constructs a polyline mesh that represents the slender parts using the screen-space shortest path analysis that can be efficiently solved using dynamic programming.  
We demonstrate the effectiveness of our approach in several in-the-wild examples.
\end{abstract}


\begin{teaserfigure}
  \includegraphics[width=\textwidth]{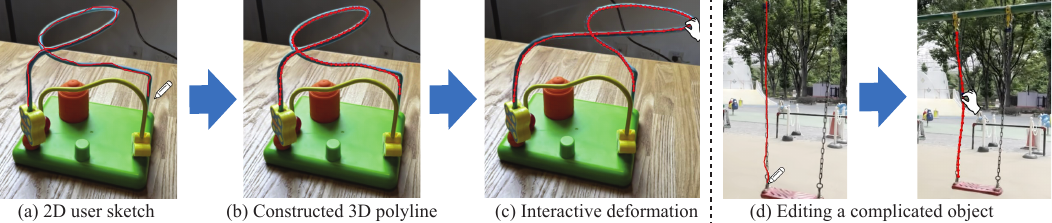}
  \caption{Given a user's sketch~(a) on a Gaussian splatting model, the system constructs a 3D polyline for elastic rod simulation~(b), where the user can deform it by pulling it~(c). It is possible to capture and edit complex shapes such as chains (d).}
  \label{fig:teaser}
\end{teaserfigure}



\maketitle

\section{Introduction}
\label{sec:intro}

Recently, the Gaussian splatting~\cite{kerbl2023_3dgs} has attracted significant attention as an efficient and high-quality novel view synthesis method.
The Gaussian splatting represents the radiance field with a set of Gaussian primitives; hence, it is possible to dynamically animate the scene by changing the properties such as the position and shape of the primitives.
Since there are numerous small Gaussian primitives in the scene, it is not practical to animate the primitives individually. 
Instead, the primitives are typically animated by embedding in the mesh deformation~\cite{gao2024_real}, tetrahedral mesh~\cite{jiang2024_vr}, skinning transformations, or spatial interpolation from points in the point-based simulation~\cite{Feng2025_gaussian}. 
Physics simulation or rigged animation is performed on such an embedded deformation space.

When constructing a deformation embedding (i.e., parametric deformation), the deformation modes need to resolve the objects' geometric detail for detailed animation. 
This is particularly true for thin or slender structures such as cloth and rods, where the neighboring geometries (e.g., body and cloth) move independently. 
The animation of the thin or slender scene typically requires discretization using a mesh that conforms to the geometry. 
Furthermore, the mesh is required to be high quality (i.e., clean topology and uniform edge length) to run a physics simulation on it.
Constructing such a \textit{simulation-ready mesh} from a model of novel view synthesis is challenging since the novel view synthesis is optimized for view interpolation and thus its geometric representation is noisy and inconsistent with the actual objects. 
In Gaussian splatting, one can possibly construct a mesh by connecting the 3D center positions of Gaussian primitives based on their proximity. 
However, this approach tends to be slow and not robust since Gaussian Splatting prioritizes visual consistency over geometric accuracy.

%
%

To address this limitation, we propose a new screen-space algorithm for constructing 3D polylines suitable for elastic rod simulations.
Our method lets the user sketch upon the slender part, then it generates the polyline mesh suitable for elastic rod simulation. 
Our method leverages dynamic programming to enable fast, interactive construction of rod meshes, and introduces a novel procedure to handle self-intersections in screen space.
The proposed approach allows slender objects to be reconstructed and edited efficiently, making it a practical tool for simulation-driven content creation.

Our contributions are summarized as
\begin{itemize}
  \item a novel screen-space algorithm for capturing slender structure from a sketch stroke from a Gaussian splatting scene, efficiently leveraging the dynamic programming,
  \item handling partially occluded objects, which is challenging for screen-space methods,
  \item several Gaussian splatting datasets for slender objects.
\end{itemize}

\section{Related Work}
\label{sec:related_work}

\paragraph{Neural Deformation \& Segmentation}
Lan et al.~\shortcite{lan2024_2d} segment the Gaussian splatting scene based on the semantic segmentation on the projected 2D image.
Qu et al.~\shortcite{qu2025_drag} introduced the method to drag the Gaussian splatting objects by specifying target positions.
However, the use of the score distillation model in moving the Gaussian primitive is computationally expensive.
Our work is based on the geometric analysis of the projected Gaussian splatting; hence, our method is lightweight and does not depend on prior knowledge.

\paragraph{Rigged Deformation}
Characters represented using Gaussian splatting are often animated using linear blend skinning (e.g., ~\cite{Kocabas_2024_CVPR}).
Ma et al.~\shortcite{ma2024_3d} combine skinning with the blendshape to construct a high-quality, animatable head model.
Our method also constructs skinning to animate the slender objects in real-time. 
While the rigging for character is typically constructed by template fitting or neural model (e.g.,~\cite{xu2020_rignet}), we construct the rigging based on the polyline extracted from the Gaussian splatting.

\paragraph{Mesh Embedding}
To animate Gaussian splatting with elastic simulation, VR-GS~\cite{jiang2024_vr} introduces a two-level embedding of Gaussian primitives inside tetrahedral meshes. 
However, it is costly to build a tetrahedral mesh. 
SuGaR~\cite{guedon2024_sugar} extracts a surface mesh from Gaussian splatting by aligning the Gaussian primitive to the surface.
To achieve large deformation, Gao et al.~\cite{gao2024_real} embed the Gaussian primitives in the faces of a surface mesh that can be adaptively subdivided.
Given a surface mesh, we can technically construct a polyline from the surface mesh using skeleton extraction (e.g.,~\cite{au2008_skeleton}), but it is still costly to construct a mesh from Gaussian splatting.

\paragraph{Meshless Embedding}
PhysGaussian~\cite{xie2024_physgaussian} and Gaussian splashing~\cite{Feng2025_gaussian} simulate deformation using point-based discretization. 
Elastic deformation using point-based discretization is suitable for bulky objects, but not suitable for slender objects such as cloth and rods.
The deformation field needs to conform to the geometry on which we run the simulation. 
The elastic rods need a polyline mesh that aligns with the slender structure.

\paragraph{Sketch-based 3D Scene Manipulation}
Our work is greatly inspired by the sketch-based modeling studies in the computer graphics community.
Specifically, 3-sweep~\cite{chen2013_3sweep} provides an interface to manipulate the image from sketch input.
Skippy~\cite{krs2017_skippy} is a sketching interface for designing 3D curves that wraps around a 3D object. 
Please refer to~\cite{liu2025_state} for the recent survey.
It is still challenging to incorporate an interactive sketch-based interface for Gaussian splatting because shape analysis is complex for point-based representation.
\begin{figure*}[h!]
    \centering
    \includegraphics[width=\linewidth]{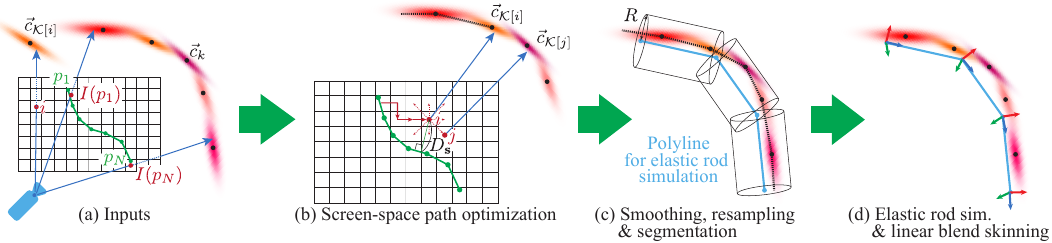}
    \caption{Workflow. (a) The user inputs a stroke $\mathbf{s}$ in a viewpoint, and the index of the Gaussian primitive for each pixel $\mathcal{K}$ is computed for that viewpoint. (b) The shortest path connecting pixels is computed by dynamic programming. (c) We smooth and resample the polyline connecting the center of Gaussian primitives to obtain a polyline for simulation. We segment the primitives inside the cylinder with radius $R$. (d) We animate the Gaussian primitives using the linear blend skinning based on the elastic rod simulation on the polyline. }
    \label{fig:placeholder}
\end{figure*}

\section{Method}
\label{sec:method}

\paragraph{Challenge}
The core of our system lies in the robust connectivity finding among the Gaussian primitives.  
Na\"ive approach finds the connectivity based on the proximity between Gaussian primitives, but it is both slow and memory-intensive. 
Furthermore, the Gaussian primitives in the slender part typically take a highly elongated shape, making it challenging to analyze proximity.
Our approach finds the connectivity in the \textit{2D screen space} instead of the \textit{3D object space} (see Fig.~\ref{fig:placeholder}).
Our key observation is that even the Gaussian primitives are not geometrically connected in the 3D space; they are connected once projected onto the screen. 
After all, the Gaussian splatting is optimized for the loss based on the projected image.

\paragraph{Image of Primitive Index}
For each pixel, we store the index of the Gaussian primitive if a primitive can be seen in the pixel. 
Multiple Gaussian primitives can be seen in a pixel since each Gaussian primitive has an opacity channel, and the final image is obtained by alpha blending.  
Thus, we store the index of the primitive that contributes the most to the pixel. 
The primitive can be efficiently found by slightly modifying the rasterization procedure of the Gaussian splatting~\cite{kerbl2023_3dgs}.
We denote the primitives index as $\mathcal{K}\left[i\right]$ that is associated with the pixel $i$.

\paragraph{User Inputs} Let the user's stroke be a sequence of coordinates on the screen $\mathbf{s}=\{p_1,p_2,\ldots,p_N\}$. 
We denote $I: \mathbb{R}^2 \rightarrow \mathbb{N}$ as the function that returns the index of the pixel given the screen coordinates. 
We assume that the stroke's first vertex and the last vertex are on the object that we want to extract. 
We denote the Gaussian primitives corresponding to the first and the last vertices as $k_{first} =\mathcal{K}\left[ I(p_1)\right]$, $ k_{last} =\mathcal{K}\left[ I(p_N)\right]$.
However, our algorithm, which we describe below, allows the points in the middle that are not precisely on top of the object we want to extract.
The user also gives the radius $R\in\mathbb{R}$ of the slender object.

\paragraph{Path Optimization}
We find the path on the screen by connecting pixels from $I(p_1)$ to $I(p_N)$.
Our method finds the path on the screen, minimizing the length of the 3D path that connects Gaussian primitives from $k_{first}$ to $k_{last}$ while following the user's input stroke. 
Given pixel $i$ and $j$ are connected, we compute the weight $w_{ij}$ as
\begin{equation}
\label{eqn:weight}
    w_{ij} = \left| \vec{c}_{\mathcal{K}\left[i\right]} - \vec{c}_{\mathcal{K}\left[j\right]} \right| + \alpha D_\mathbf{s}\left(I^{-1}(i)\right)^2,
\end{equation}
where $\vec{c}_k\in\mathbb{R}^3$ is the center of the Gaussian primitive $k$, $D_\mathbf{s}$ is the minimum distance to the polyline $\mathbf{s}$ and a point on the screen, and $I^{-1}: \mathbb{N}\rightarrow \mathbb{R}^2$ is the function that returns the pixel center given the pixel index. 
Note that, in \eqref{eqn:weight}, the first term evaluates the length of the polyline in 3D and the second term evaluates the distance to the input stroke while $\alpha\in\mathbb{R}$ is a user-defined parameter controlling the tolerance of the deviation of the path from the stroke ($\alpha=1$ in this paper). 
A pixel on the screen can connect to adjacent eight pixels that share an edge or a vertex. 
Dynamic programming efficiently finds the path connecting adjacent pixels that minimizes the sum of the weights~\eqref{eqn:weight}.
We are inspired by the use of dynamic programming in the sketch snapping~\cite{su2014_ez}.

\begin{figure}[b!]
    \centering
    \includegraphics[width=\linewidth]{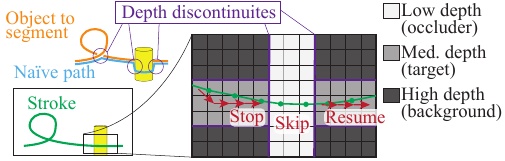}
    \caption{Left: Naively connecting the shortest path on the screen may result in discontinuities when there are occlusions. Right: to overcome this issue, we do not connect pixels with a large depth gap. In addition, we skip the vertices in the stroke if there is a depth gap.}
    \label{fig:occlusion}
\end{figure}

\paragraph{Occlusion} 
The algorithm we described so far works well if all the slender parts are visible, i.e., not occluded by themselves or other objects.
However, as shown in Fig.~\ref{fig:occlusion}-left, the shortest path may skip the looped part at the self-intersection or the path may jump to the occluding object.
In both cases, the shortest path contains a discontinuous depth change between adjacent pixels. 
Thus, in the dynamic programming, we exclude such pixels from the candidates when the distance is greater than the $3R$.

By eliminating the discontinuous adjacent pixels from the candidate, our algorithm may stop where there is no candidate (see Fig.~\ref{fig:occlusion}-right). 
To skip the occluded part, we check the distances between the vertex of the stroke and the path connecting pixels and flag \textit{covered} if it is less than a threshold (10 pixels here).
Then, we resume the path finding from the not-\textit{covered} vertex of the stroke with the smallest index whose depth value is similar to the pixel where the connection is lost.
Note that our occlusion handling has some heuristics and sometimes fails. 
However, since our algorithm is very fast, the user can always try again by sketching a more accurate stroke or sketching from a viewpoint with less occlusion.

%

\paragraph{Post Process}
After computing the shortest path connecting pixels on the screen, we construct a 3D polyline mesh by connecting the centers of the Gaussian primitives that are associated with the pixels.
Since the centers of the Gaussian primitives and the centers of the pixels are not aligned perfectly, the initial 3D polyline is jagged. 
We apply a few steps of Laplacian smoothing, and then we resample the polyline with a user-specified edge length. 
Once the polyline mesh is computed, we segment each Gaussian primitive if its center's shortest distance to the polyline is less than $R$.

Finally, we animate Gaussian primitives using the linear blend skinning.
The skinning weight is computed by finding the nearest point on the polyline to the center of the Gaussian. 
For the physics simulation of elastic rods, we use the discrete elastic rod where the rotation frames are locally updated~\cite{bergou2010_discrete}. 
The frame on a vertex is computed by averaging ones on adjacent edges.

\section{Results}
\label{sec:results}

\begin{figure*}[h!]
    \centering
    \includegraphics[width=\linewidth]{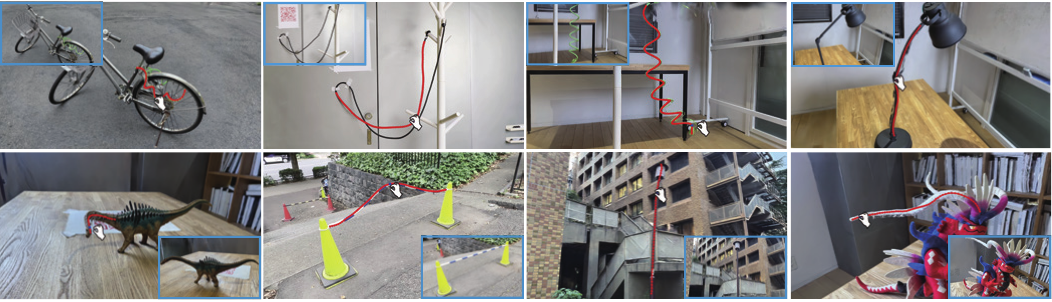}
    \caption{Examples of manipulating slender objects using our system. Each pair images shows the undeformed object (inset) and resulting real-time deformation by pulling the 3D polyline.}
    \label{fig:results}
\end{figure*}

\paragraph{Implementation detail} 
The dynamic programming and the elastic rod simulation are implemented in C++ and Rust, respectively, while the interface is implemented using Python~\footnote{The code is available at \url{https://github.com/haato-w/sketch-rod-gs}}. 
Because our method focuses on polyline mesh construction for skinny object, We collected in-the-wild 10 datasets for 3D Gaussian splatting that has slender objects (e.g., chain, tube and pole). 
The training images are Full HD, taken by author's iPhone 15 Pro.
All the experiments are conducted with an image resolution of 960 x 540 on a machine with a GeForce RTX 3090 GPU and an Intel Core i9 CPU. 
Fig.~\ref{fig:results} shows examples of the editing results using our system.
For all the examples, our algorithm roughly takes 0.5 second for processing the sketch and $50$ milliseconds for animation and rendering. 
Please refer to the accompanying video for the animation results.

\begin{figure}[htbp!]
    \centering
    \includegraphics[width=\linewidth]{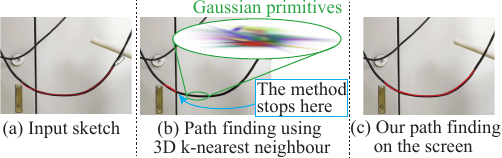}
    \caption{Comparison against a baseline using 3D k-nearest neighbor path finding. Given a stroke (a), the na\"ive method stops working in the middle due to noisy, highly elongated Gaussian primitives (b). On the other hand, our screen space approach finds the path covering the entire stroke (c). }
    \label{fig:comparison}
\end{figure}

\paragraph{Comparison to the baseline} 
Fig.~\ref{fig:comparison} shows the comparison against the na\"ive method, where the connection between the primitives is optimized with dynamic programming with the same  weight~\eqref{eqn:weight}, but the candidates are selected from the eight nearest neighbors of the primitives' center based on 3D Euclidean distance. 
While the na\"ive method stops computing the path in the middle due to the highly elongated anisotropic Gaussian primitives.
The na\"ive method, implemented in Python, takes 13 seconds  to compute until it stops, excluding the time to construct the acceleration structure.

\vskip 10mm

\section{Conclusions \& Future Work}

In this work, we present an interactive system to construct a polyline mesh for elastic rod simulation from a Gaussian splatting model.
We demonstrate the efficiency and robustness of our screen space algorithm that leverages dynamic programming.

However, when a smooth polyline bends and forms a cusp, the individual needle-like elongated Gaussian primitives might be visible as an artifact. 
This artifact can be prevented by subdividing the Gaussian primitives beforehand~\cite{gao2024_real}.
Currently, the user needs to set the radius of the slender object for the segmentation. 
It is left as future work to recognize the shape of the slender part as a generalized cylinder~\cite{chen2013_3sweep}  by extracting the medial axis and cross-section shape.
%
Finally, an interesting future research avenue is to construct high-quality surface meshes capable of thin-shell or cloth simulation from screen-space analysis. 

\begin{acks}
This work was financially supported by I. Meisters inc.
\end{acks}

\bibliographystyle{ACM-Reference-Format}
\bibliography{reference}

\end{document}